\documentclass[12pt]{iopart}
\usepackage{iopams}
\usepackage{epsfig}
\input{epsf}

\begin{document}
\title{Cosmological Landscape and Euclidean Quantum Gravity}
%\date{}
\author{A O Barvinsky$^{1,2}$ and A Yu Kamenshchik$^{3,4}$}
\address{$^{1}$\em Theory Department, Lebedev
Physics Institute, Leninsky Prospect 53, 119991 Moscow, Russia\\
$^{2}$Sektion Physik, LMU, Theresienstr. 37, Munich, Germany\\
$^{3}$Dipartimento di Fisica and INFN, Via Irnerio 46, 40126
Bologna, Italy\\
$^{4}$L.D.Landau Institute for Theoretical Physics, Kosygin Str. 2,
119334 Moscow, Russia} \eads{\mailto{barvin@lpi.ru},
\mailto{kamenshchik@bo.infn.it}}
\begin{abstract}
Quantum creation of the universe is described by the {\em density
matrix} defined by the Euclidean path integral. This yields an
ensemble of universes --- a cosmological landscape --- in a mixed
quasi-thermal state which is shown to be dynamically more preferable
than the pure quantum state of the Hartle-Hawking type. The latter
is suppressed by the infinitely large positive action of its
instanton, generated by the conformal anomaly of quantum matter. The
Hartle-Hawking instantons can be regarded as posing initial
conditions for Starobinsky solutions of the anomaly driven deSitter
expansion, which are thus dynamically eliminated by infrared effects
of quantum gravity. The resulting landscape of hot universes treated
within the cosmological bootstrap (the self-consistent back reaction
of quantum matter) turns out to be limited to a bounded range of the
cosmological constant, which rules out a well-known infrared
catastrophe of the vanishing cosmological constant and suggests an
ultimate solution to the problem of unboundedness of the
cosmological action in Euclidean quantum gravity.
\end{abstract}
\submitto{\JPA}
\section{Introduction}
The ideas of quantum cosmology \cite{Bryce,VilNB} and Euclidean
quantum gravity \cite{HH,H} are again attracting attention. One of
the reasons is the fact that the landscape of string vacua is too
big \cite{landscape} to hope that a reasonable selection mechanism
can be successfully worked out
within string theory itself. Thus, it is expected that other methods
have to be invoked, at least some of them appealing to the
construction of the cosmological wavefunction
\cite{OoguriVafaVerlinde,Tye1,Tye23,Brustein}.
This quantum state arises as a result of quantum tunneling from the
classically forbidden state of the gravitational field.
The
Hartle-Hawking wave function of the Universe \cite{HH,H}
describes nucleation of the de Sitter Universe from the Euclidean
4-dimensional hemisphere,$
    \Psi_{\rm HH}\sim \exp(-S_{\rm E})
    =\exp(3\pi /2G\Lambda)$ and has a negative action which diverges
    to $-\infty$ for the
cosmological constant $\Lambda\to 0$. This implies a well-known
infrared catastrophe of small cosmological constant --- a vanishing
$\Lambda$ is infinitely more probable than any positive one.
Alternative tunneling proposals for the wave function of the
universe in the form of Linde \cite{Linde} or Vilenkin
\cite{Vilenkin} give preference to big values of $\Lambda$, which
opens the possibility for conclusions opposite to the Hartle-Hawking
case. In particular, the inclusion of one-loop effects allows one to
shift most probable values of the effective cosmological constant
from zero to a narrow highly peaked range compatible with the
requirements of inflation \cite{scale}.

\begin{figure}[h]
\centerline{\epsfxsize 5cm \epsfbox{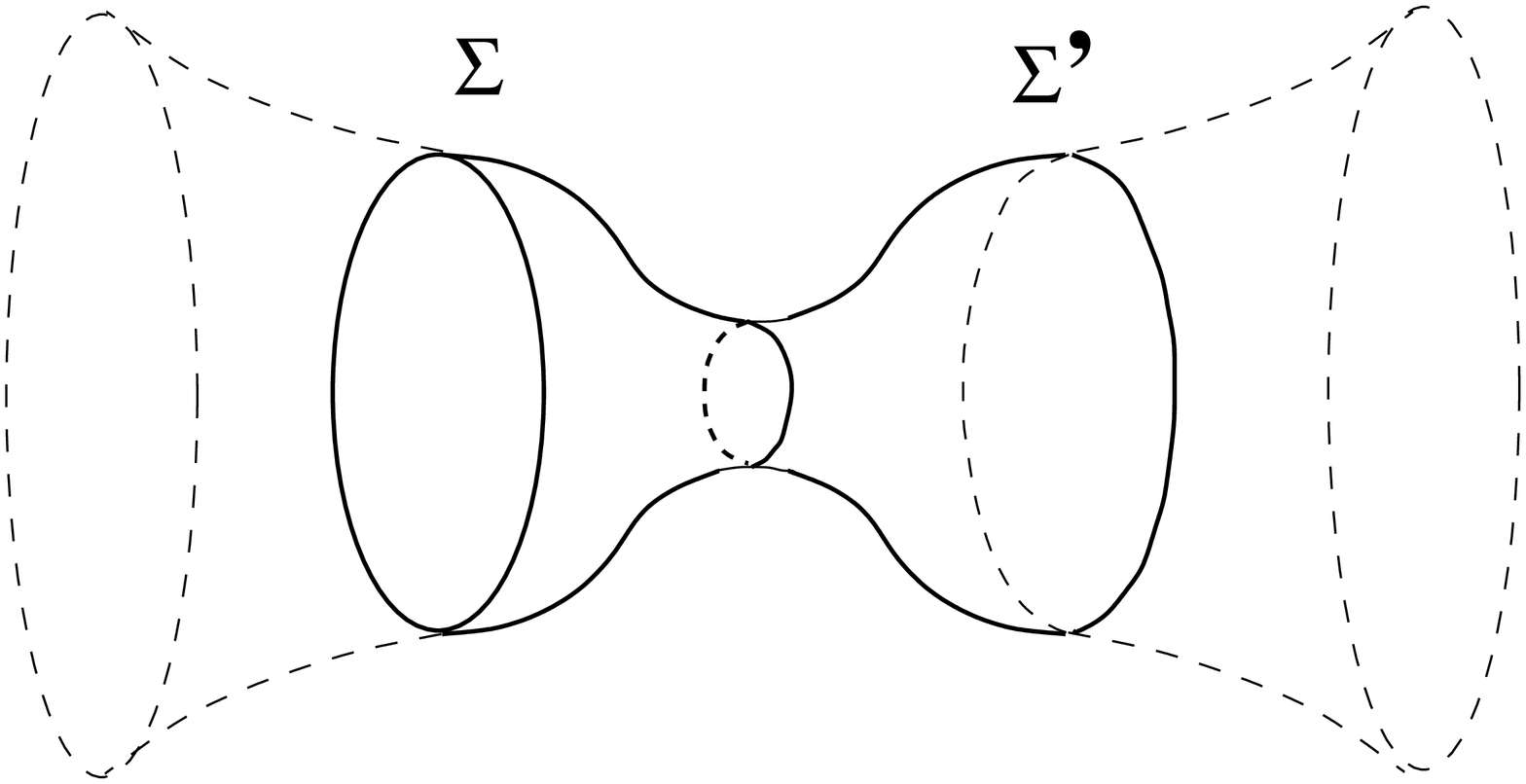}} \caption{\small
Picture of instanton representing the density matrix. Dashed lines
depict the Lorentzian Universe nucleating from the instanton at the
minimal surfaces $\Sigma$ and $\Sigma'$. \label{Fig.1}}
\end{figure}

\begin{figure}[h]
\centerline{\epsfxsize 4.3cm \epsfbox{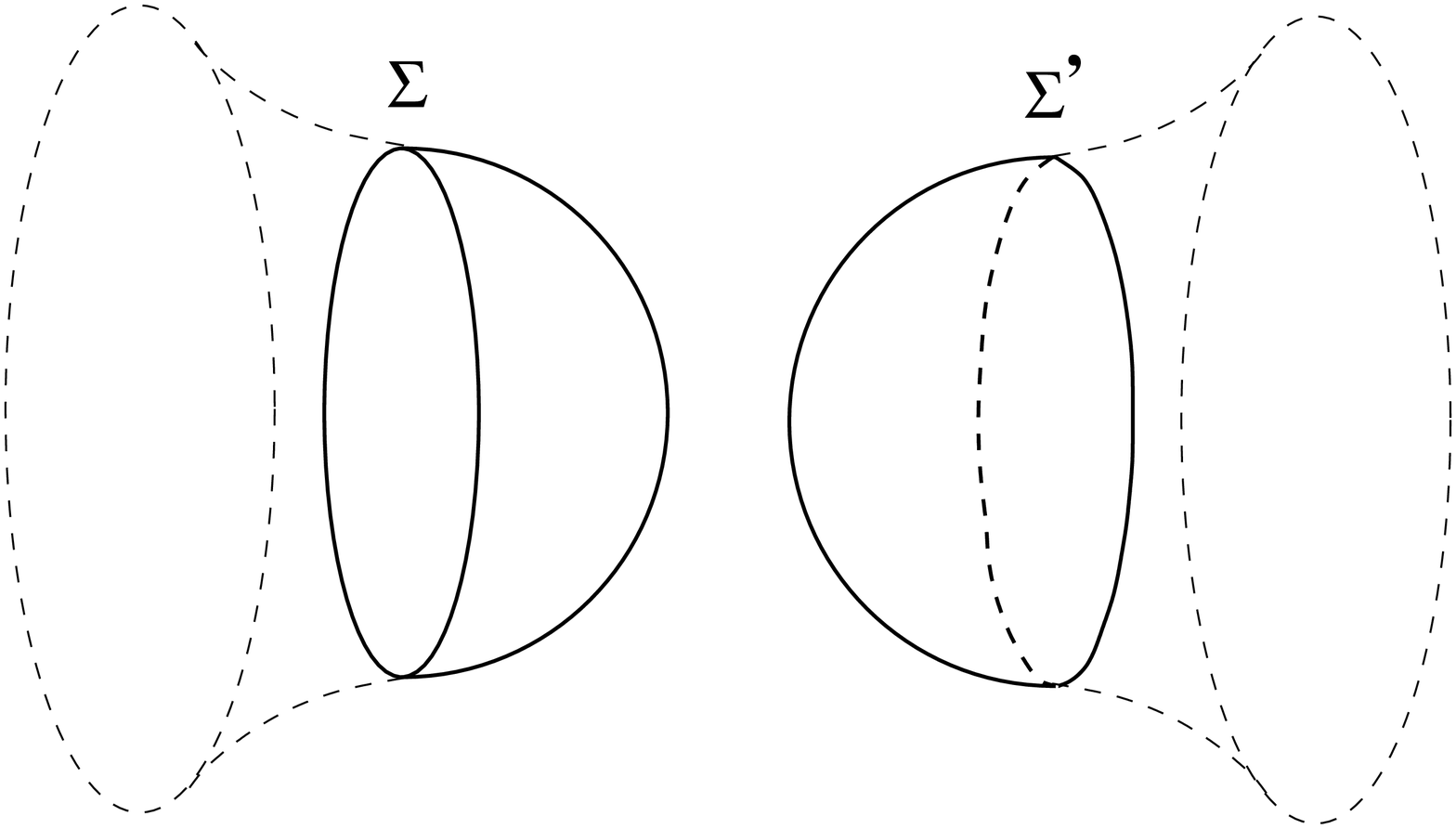}} \caption{\small
Density matrix of the pure Hartle-Hawking state represented by the
union of two vacuum instantons. \label{Fig.2}}
\end{figure}

In this work we study the Hartle-Hawking prescription of the
Euclidean path integration taking into account
essentially {\em nonlocal} quantum
effects mediated by nonlocal terms of non-vacuum nature.
The core of our suggestion is a simple observation that the presence
of radiation implies a statistical ensemble described by the density
matrix, rather than a pure state assumed in \cite{Tye1,Tye23}.
Density matrix in Euclidean quantum gravity \cite{Page},
$\rho[\,\varphi,\varphi']$, originates from an instanton with two
disjoint boundaries $\Sigma$ and $\Sigma'$ associated  respectively
with its two entries,  see
Fig.\ref{Fig.1}.  Note that mixed nature of the quantum state is fundamental and the
origin  of impurity is not caused by coarse graining or tracing out
environmental degrees of freedom.
\\
In contrast, the pure density matrix of the Hartle-Hawking state
corresponds to the situation when the instanton bridge between
$\Sigma$ and $\Sigma'$ is broken, so that topologically the
instanton is a union of two disjoint hemispheres. Each of the
half-instantons smoothly closes up at its pole which is a regular
internal point of the Euclidean spacetime ball, see Fig.\ref{Fig.2}
--- a picture illustrating the factorization of
$\hat\rho=|\Psi_{\rm HH}\rangle\langle\Psi_{\rm HH}|$.
\\

When calculated in the saddle-point approximation the density matrix
automatically gives rise to radiation whose thermal fluctuations
destroy the Hartle-Hawking instanton. Namely, the radiation stress
tensor prevents the half-instantons of the above type from closing
and, thus, forces the tubular structure on the full instanton
supporting the thermodynamical nature of the physical state. The
existence of radiation, in its turn, naturally follows from the
partition function of this state. The partition function originates
from integrating out the field $\varphi$ in the coincidence limit
$\varphi'=\varphi$. 
(This procedure is equivalent to taking the trace
of the  operator $\exp(-\beta H)$ inherent in the calculation of the
partition function for a system with the Hamiltonian $H$.)
This corresponds to the identification of
$\Sigma'$ and $\Sigma$, so that the underlying instanton acquires
toroidal topology. Its points are labeled by
the periodically identified Euclidean time, a period being related
to the inverse temperature of the quasi-equilibrium radiation. The
back reaction of this radiation supports the instanton geometry in
which this radiation exists, and we derive the equation which makes
this bootstrap consistent.

We show that for the radiation of conformally-invariant fields the
analytical and numerical solution of the bootstrap equations yields
a set of instantons --- a landscape --- only in the bounded range of
$\Lambda$,
    \begin{eqnarray}
    \Lambda_{\rm min}<\Lambda<\Lambda_{\rm max}.            \label{2}
    \end{eqnarray}
This set consists of the countable sequence of one-parameter
families of instantons which we call garlands, labeled by the number
$k=1,2,3,...$ of their elementary links. Each of these families
spans a continuous subset, $\Lambda_{\rm
min}^{(k)}<\Lambda\leq\Lambda_{\rm max}^{(k)}$, belonging to
(\ref{2}). These subsets of monotonically decreasing lengths
$\Lambda_{\rm max}^{(k)}-\Lambda_{\rm min}^{(k)}\sim 1/k^4$ do not
overlap and their sequence has an accumulation point at the upper
boundary $\Lambda_{\rm max}$ of the range (\ref{2}). Each of the
instanton families at its upper boundary $\Lambda_{\rm max}^{(k)}$
saturates with the static Einstein Universe filled by a hot
equilibrium radiation with the temperature $T_{(k)}\sim m_P/\ln
k^2$, $k\gg1$, and having the {\em negative} decreasing with $k$
action $\Gamma_0^{(k)}\sim -\ln^3k^2/k^2$. Remarkably, all values of
$\Lambda$ below the range (\ref{2}), $\Lambda<\Lambda_{\rm min}$,
are completely ruled out either because of the absence of instanton
solutions or because of their {\em infinitely large positive}
action.

The details of our approach have recently been presented in
\cite{we}, while here we discuss its basic points and, in
particular, establish its relation to the well-known Starobinsky
model \cite{Starob} of the self-consistent deSitter expansion driven
by the conformal anomaly of quantum fields.

\section{The effective action, generalized Friedmann equations and bootstrap}
As shown in \cite{we} the effective action of the cosmological model
with a generic spatially closed FRW metric $ds^2 = N^2(\tau) d\tau^2
+a^2(\tau)d^2\Omega^{(3)}=a^2(d\eta^2 + d^2\Omega^{(3)})$ sourced by
{\em conformal} quantum matter has the form
\begin{eqnarray}
    &&\Gamma[\,a(\tau),N(\tau)\,]=
    2 \int_{\tau_-}^{\tau_+} d\tau\Big(\!-\frac{a\dot{a}^2}N
    -Na+N H^2 a^3\Big)\nonumber\\
    &&\qquad\qquad\qquad\quad
    +2B \int_{\tau_-}^{\tau_+}
    d\tau \Big(\frac{\dot{a}^2}{Na}
    -\frac16\,\frac{\dot{a}^4}{N^3 a}\Big)\nonumber\\
    &&\qquad\qquad\qquad\quad
    + B \int_{\tau_-}^{\tau_+}
    d\tau\,N/a+F\Big(2\int_{\tau_-}^{\tau_+}
    d\tau\,N/a\Big)\, ,                      \label{Gamma}
    \end{eqnarray}
where $a(\tau)$ is the cosmological radius, $N(\tau)$ is the lapse
function, $H^2 = \Lambda/3$ and integration runs between two turning
points at $\tau_{\pm}$, $\dot a(\tau_\pm)=0$. Here the first line is
the classical part, the second line is the contribution of the
conformal transformation to the metric of the static instanton
$d\bar{s}^2 = d\eta^2 + d^2\Omega^{(3)}$ ($\eta$ is the conformal
time) and the last line is the one-loop action on this static
instanton. The conformal contribution $\Gamma_{\rm
1-loop}[g]-\Gamma_{\rm 1-loop}[\bar g]$ is determined by the
coefficients of $\Box R$, the Gauss-Bonnet invariant $E =
R_{\mu\nu\alpha\gamma}^2 -4R_{\mu\nu}^2 + R^2$ and Weyl tensor term
in the conformal anomaly
    $g_{\mu\nu}\delta
    \Gamma_{\rm 1-loop}/\delta g_{\mu\nu} =
    g^{1/2}
    (\alpha \Box R +
    \beta E + \gamma C_{\mu\nu\alpha\beta}^2)/4(4\pi)^2 $.
Specifically this contribution can be obtained by the technique of
\cite{TseytlinconfBMZ}; it contains higher-derivative terms $\sim
\ddot a^2$ which produce ghost instabilities in solutions of
effective equations. However, such terms are proportional to the
coefficient $\alpha$ which can be put to zero by adding the
following finite {\em local} counterterm
    \begin{eqnarray}
    \Gamma_{R}[g]
    =\Gamma_{\rm 1-loop}[g]
    +\frac1{2(4\pi)^2}
    \frac\alpha{12}\int d^4x\,
    g^{1/2}R^2(g).                   \label{renormalization}
    \end{eqnarray}
This ghost-avoidance renormalization is justified by the requirement
of consistency of the theory at the quantum level. The contribution
$\Gamma_{R}[g]-\Gamma_{R}[\bar g]$ to the {\em renormalized}
action then gives the second line of (\ref{Gamma}) with
$B=3\beta/4$.

The static instanton with a period $\eta_0$ playing the role of
inverse temperature contributes $\Gamma_{\rm 1-loop}[\bar g]
=E_0\,\eta_0+F(\eta_0)$, where the vacuum energy $E_0$ and free
energy $F(\eta_0)$ are the typical boson and fermion sums over field
oscillators with energies $\omega$ on a unit 3-sphere
    %\begin{equation}
    $E_0=\pm\sum_{\omega}
    \omega/2\,,\,\,\,F(\eta_0)=\pm\sum_{\omega}
    \ln\big(1\mp e^{-\omega\eta_0}\big)$.
    %\end{equation}
The renormalization (\ref{renormalization}) which should be applied
also to $\Gamma_{\rm 1-loop}[\bar g]$ modifies $E_0$, so that
$\Gamma_{R}[\bar g] =C_0\,\eta_0+F(\eta_0)$, $C_0\equiv
E_0+3\alpha/16$. This gives the third line of Eq.(\ref{Gamma}) with
$C_0=B/2$. This universal relation between $C_0$ and $B=3\beta/4$
follows from the known anomaly coefficients  \cite{confanomaly} and
the Casimir energy in a static universe \cite{E_0} for scalar, Weyl
spinor and vector fields.

The Euclidean Friedmann equation looks now as
\begin{eqnarray}
    &&\frac{\dot{a}^2}{a^2}
    +B \left(\frac12\,\frac{\dot{a}^4}{a^4}
    -\frac{\dot{a}^2}{a^4}\right) =
    \frac{1}{a^2} - H^2 -\frac{C}{ a^4},     \label{efeq}\\
    %\end{eqnarray}
    %\begin{eqnarray}
    &&C = B/2 +F'(\eta_0),\,\,
    \eta_0 = 2\int_{\tau_-}^{\tau_+}
    d\tau/a(\tau).                       \label{bootstrap}
    \end{eqnarray}
The contribution of the nonlocal $F(\eta_0)$ in (\ref{Gamma})
reduces here to the radiation {\em constant} $C$ as a {\em nonlocal
functional} of $a(\tau)$, determined by the {\em bootstrap} equation
(\ref{bootstrap}), $F'(\eta_0)\equiv dF(\eta_0)/d\eta_0>0$ being the
energy of a hot gas of particles, which adds to their vacuum energy
$B/2$.

\begin{figure}[h]
\centerline{\epsfxsize 8.5cm \epsfbox{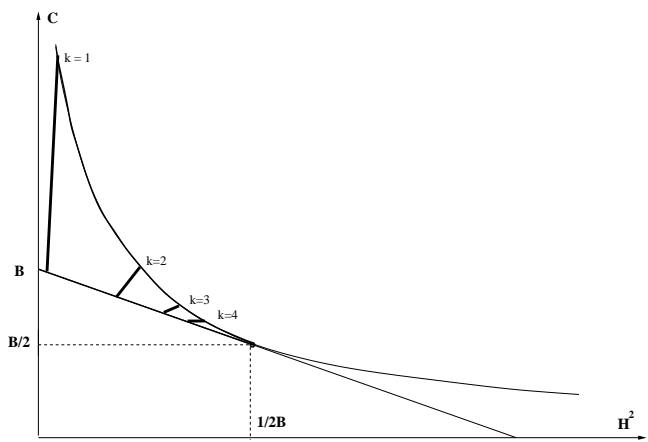}} \caption{\small
Instanton domain in the $(H^2,C)$-plane. Garland families are shown
for $k=1,2,3,4$. Their sequence accumulates at the critical point
$(1/2B,B/2)$.
 \label{Fig.4}}
\end{figure}
Periodic instanton solutions of Eqs.(\ref{efeq})-(\ref{bootstrap})
exist only inside the curvilinear wedge of $(H^2,C)$-plane between
bold segments of the upper hyperbolic boundary and the lower
straight line boundary of Fig.\ref{Fig.4},
    \begin{equation}
    4CH^2\leq 1,\,\,\,C \geq B-B^2 H^2,
    \,\,\,B H^2\leq 1/2.             \label{restriction1}
    \end{equation}
Below this domain the solutions are either complex and aperiodic or
suppressed by {\em infinite positive} Euclidean action. Indeed,
a smooth Hartle-Hawking instanton
with $a_-=0$  yields  $\eta_0\to\infty$ in view of
(\ref{bootstrap}), so that $F(\eta_0)\sim F'(\eta_0)\to 0$.
Therefore, its on-shell action
    \begin{equation}
    \Gamma_0= F(\eta_0)\!-\!\eta_0 F'(\eta_0)
    +4\!\int_{a_-}^{a_+}\!
    \frac{da \dot{a}}{a}\Big(B-a^2
    -\frac{B\dot{a}^2}{3}\Big)              \label{action-instanton}
    \end{equation}
due to $B>0$ diverges to $+\infty$ at $a_-=0$ and completely rules
out pure-state instantons \cite{we}. For the instanton garlands,
obtained by glueing together into a torus $k$ copies of a simple
instanton \cite{Halliwell} the formalism is the same as above except
that the conformal time in (\ref{bootstrap}) and the integral term
of (\ref{action-instanton}) should be multiplied by $k$. As in the
case of $k=1$, garland families interpolate between the lower and
upper boundaries of (\ref{restriction1}). They exist for all $k$,
$1\leq k\leq\infty$, and their infinite sequence accumulates at the
critical point $C=B/2$, $H^2=1/2B$, with the on-shell action
$\Gamma_0^{(k)}\simeq -B\,\ln^3 k^2/4k^2\pi^2$ which, in contrast to
the tree-level garlands of \cite{Halliwell} is not additive in $k$.

\section{Conclusions: Euclidean quantum gravity and the Starobinsky model}
Elimination of the Hartle-Hawking vacuum instantons implies also
ruling out well-known solutions in the Starobinsky model of the
anomaly (and vacuum energy) driven deSitter expansion \cite{Starob}.
Such solutions (generalized to the case of a nonzero ``bare"
cosmological constant $3H^2$) can be obtained by the the usual Wick
rotation from the solutions of (\ref{efeq}) with the thermal
contribution switched off. The corresponding Euclidean Friedmann
equation reads
\begin{equation}
\frac{\dot{a}^2}{a^2} - \frac{1}{a^2} +
\frac{B}{2}\left(\,\frac{\dot{a}^2}{a^2} - \frac{1}{a^2}\,
\right)^2+ H^2=0. \label{star}
 \end{equation}
and has a generic solution of the form $a = \sin h\tau/h$,
$1/a^2-\dot a^2/a^2=h^2$, with the following two values of the
Hubble parameter $h=h_\pm$,
%\begin{equation}
$h^2_\pm = (1/B)(1\pm \sqrt{1 - 2BH^2})$.
%\label{star1}
%\end{equation}
For $H = 0$ this is exactly the Euclidean version of the
Starobinsky's solution \cite{Starob} with $h_+^2 = 2/B$. For larger
$H^2<1/2B$ we have two families of exactly deSitter instantons which
can be regarded as initial conditions for the ``generalized"
solutions of Starobinsky. However, all of them are ruled out by
their infinite positive Euclidean action.

What remains is a quasi-thermal ensemble of non-vacuum models in the
bounded cosmological constant range (\ref{2}) with $\Lambda_{\rm
min}>0$ and $\Lambda_{\rm max}=3/2B$ and with a finite value of
their effective Euclidean action. This implies the elimination of
the infrared catastrophe of $\Lambda\to 0$ and the ultimate solution
to the problem of unboundedness of the {\em on-shell} cosmological
action in Euclidean quantum gravity. As a byproduct, this suggests
also strong constraints on the cosmological constant apparently
applicable to the resolution of the cosmological constant problem.

\ack{A.K. is grateful to A.A Starobinsky and A. Vilenkin for useful
discussions. A.B. is indebted to C.Deffayet for stimulating
discussions and the hospitality of IAP in Paris. A.B. was supported
by the RFBR grant 05-02-17661 and the LSS 4401.2006.2 and SFB 375
grants. A.K. was partially supported by the RFBR grant 05-02-17450
and the grant LSS 1757.2006.2.

\end{document}